\documentclass[aps,prl,twocolumn,groupedaddress]{revtex4}

\usepackage{graphicx}
\usepackage{xspace}
\usepackage{booktabs}
\usepackage{psfig}
\usepackage{amsmath}

\begin{document}

\title{Coherent Light Scattering from Semicontinuous Silver Nanoshells near the Percolation Threshold}

\author{C. A. Rohde}
\author{K. Hasegawa}
\author{M. Deutsch}

\affiliation{Oregon Center for Optics and Department of Physics, University of Oregon, Eugene, OR 97403}

\date{\today}

\begin{abstract}

We report on measurements of visible extinction spectra of semicontinuous silver nanoshells grown on colloidal silica spheres. We find that thin, fractal shells below the percolation threshold exhibit geometrically tunable plasmon resonances. A modified scaling theory approach is used to model the dielectric response of such shells, which is then utilized to obtain the extinction cross section in a retarded Mie scattering formalism. We show that such spherical resonators support unique plasmon dynamics: in the visible there is a new regime of coherently driven cluster-localized plasmons, while crossover to homogeneous response in the infrared predicts a delocalized shell plasmon.
\end{abstract}


\maketitle

Noble-metal nano-scale shells are comprised of thin gold or silver films surrounding sub-micron dielectric cores. In recent years nanometer-scale metal particles have been the focus of extensive studies, owing mainly to large enhancements of surface-induced electric fields at the plasma resonance of the nanoparticles \cite{moskovits}. This unique optical response is well exemplified in the nanoshells' extinction spectra, which are governed by a geometrically tunable plasmon resonance. When the core diameter is in the sub-micrometer range, the optical response of the composite particle is tunable over the entire visible and near infrared spectrum \cite{halas1}. This constitutes a powerful tool for custom-designing Raman \cite{halas2} and surface-plasmon-based ultra sensitive optical sensors \cite{shultz}.

The strong optical resonance observed in spherical nanoshells is attributed to a delocalized shell plasmon, typically well-modelled by classical linear response to electromagnetic (EM) plane-wave scattering \cite{halas3}. This was formulated by Aden and Kerker (AK) \cite{aden} as an extension to classical Mie scattering theory. More refined theories utilizing ab-initio quantum mechanical modelling of metallic nanoshells have also verified this approach \cite{nord1}.

The existence of a spherical--shell plasmon relies on a symmetric and uniform shell morphology, and is typically observed in smooth, continuous nanoshells. It has been previously shown that extinction spectra of incomplete, highly fragmented nanoshells do not exhibit geometrical resonances \cite{AvB}, but are dominated instead by absorption resonances of the metal clusters on the spheres \cite{davidov}.

In this Letter we present light scattering experiments of silica spheres coated with discontinuous silver nanoshells. We distinguish two shell morphologies with differing optical signatures. For thin, two-dimensional (2D) fractal shells, the optical response of the metal is best described within the framework of percolation theory. While these shells support mainly localized plasmon modes and are electrically insulating, distinct geometrical resonances are clearly present, due to coherent optical driving of cluster-localized plasmons. For thick (3D), conducting and porous shells classical, Maxwell Garnett (MG) theory is used instead to obtain the effective dielectric response of the shell. In both cases, although the shells are locally highly irregular, observed fine resonances are still well modelled by the AK solution to Maxwell's equations for uniform core/shell systems.

Colloidal silica spheres (0.3-1$\mu$m diameter, 2$\%$ polydisperse) were coated with nanocrystalline silver shells using a Tollen's reagent method to form dielectric core--metal shell particles \cite{rohde}. Aqueous ammoniacal silver nitrate was mixed with an aldehyde reducing agent such as glucose, leading to rapid precipitation of silver atoms. At 3$^{\circ}$C the reaction kinetics are dramatically slowed, resulting in controlled growth of nanocrystalline silver, only at the spheres' surfaces. Typical reactions last 20-30min. Repeated centrifugation and sonication are utilized to eliminate residual silver nanocrystals from solution. Despite a lack of surface functionalization, the silver adheres strongly to the spheres and does not separate during this process. More than 70 such batches were synthesized.

\begin{figure}[b]
    \includegraphics[width=8.5cm]{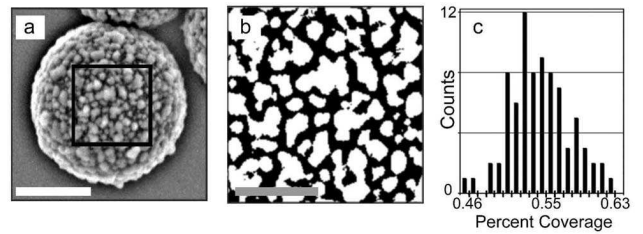}
    \caption{
        (a) Scanning electron micrograph of 1$\mu$m diameter silica sphere coated with 20nm--thick silver shell.
        Scale bar is 500nm.
    (b) Digitally processed (see text) magnified view of region inside the square in (a). Scale bar is 200nm. Silver
    appears white and dielectric inclusions are black.
    (c) Histogram of $p$, generated from 80 images similar to (b).
  }
    \label{fig:sem}
\end{figure}

Transmission electron micrographs (TEM) of more than 200 coated spheres per data set are used to determine average shell thickness and polydispersity. For thin coatings ($\sim$20 - 30nm) ensemble polydispersities can remain as low as 2$\%$, indicating highly uniform deposition. High resolution TEM analyses of isolated metal clusters of $\sim 100$nm mean size indicate the latter consist of aggregated nanocrystals 5 - 10nm in diameter.

Further characterization of the shells' morphologies utilizes high-resolution scanning electron micrographs (SEM). These reveal the non-contiguous nature of the thin shells, which we characterize in terms of their metal filling fraction, $p$, and a fractal dimension, $D_{f}$. SEM images as in Fig.\ref{fig:sem}(a) are first processed with a high pass fast Fourier transform filter to eliminate nonuniform background intensity variations across the sphere surface. Sphere curvature and shadowing effects are minimized by utilizing only a small portion of the exposed surfaces close to nadir, shown enclosed in the box in Fig.\ref{fig:sem}(a). The size of this area must be small enough to minimize shadowing effects, while still sufficiently large to provide statistical integrity. With a resolution-limited pixel size of $\simeq 2$nm, for coated spheres of radius $R$ a box size of $(R^2/4) \simeq 200^2$ pixels satisfies this condition. Linear contrast enhancement and thresholding of the cropped section provide the binary image in Fig.\ref{fig:sem}(b). Visual inspection of such images suggests a strong resemblance to percolating planar metal films deposited under ultra-high vacuum, where both linear and nonlinear optical characteristics have been previously addressed \cite{yagil92, shalaev1}.

We measure the fractal dimension of thin films such as in Fig.\ref{fig:sem}(b) by implementing a box-counting method \cite{douketis} over two orders of magnitude, and find $D_{f}=1.72 \pm 0.06$, indicating diffusion-limited aggregation in 2D \cite{jensen}. Since this is below the fractal dimension of 2D percolating clusters, $D_p = 1.89$ \cite{isichenko}, we conclude these shells lie slightly below the percolation threshold. We have verified this by measuring the average filling fraction $\overline{p}$ of thin shells utilizing the processed SEM images. For an ensemble of $\sim$100 spheres we find $\overline{p}=0.55$, shown in Fig.\ref{fig:sem}(c). Comparing to the experimental percolation threshold, $p_c$ = 0.68 \cite{yagil92}, our thin shells have $p \lesssim p_c$.

This is supported by measurements of the bulk resistance of silver-coated spheres. A dense aggregate of spheres is formed by drying several concentrated drops of the sphere suspension between two gold electrodes separated by $\sim$100$\mu$m. For shells 20 - 30nm thick we measure a resistance $R$$\sim$20M$\Omega$, while shells of thickness $\geq$70nm exhibit a resistance of $R$$\simeq$1$\Omega$. These are typical values, also obtained previously for the sheet resistance of planar thin metal films near the percolation threshold \cite{yagil92}.

\begin{figure}[t]
   \includegraphics[width=8.5cm]{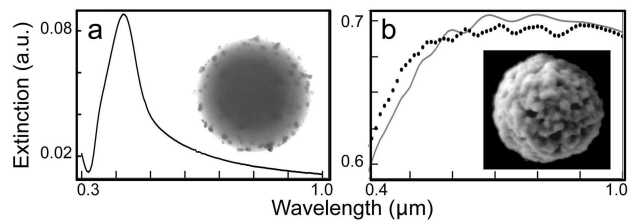}
        \caption{(a) Extinction of sparsely coated 1$\mu$m diameter spheres (TEM inset) showing a dipolar plasmon resonance at 420nm. (b) Measured extinction (dots) and fitted AK model using MG theory (line) for 70nm-thick shells (SEM inset).}
\label{fig:sparse}
\end{figure}

To optically characterize our core/shell composites we measured the extinction of freshly prepared, dilute aqueous suspensions using a UV/vis spectrometer. Spectra of sparsely coated spheres as in Fig.\ref{fig:sparse}(a) exhibit only a single peak centered near 420nm. This is a signature of the excitation of the dipolar plasmon eigenmodes of individual, non-aggregated silver nanocrystals nucleated on the spheres \cite{davidov}. Due to the dilute metal surface-coverage, these modes do not couple to the weak Mie resonances of the dielectric spheres, and are therefore independent of sphere size. At the opposite extreme are densely coated spheres as in Fig.\ref{fig:sparse}(b), exhibiting multipolar core/shell resonances \cite{halas3}. In this limit the extinction is well-modelled by applying MG theory within the AK framework, as discussed in a later section.

\begin{figure}[t]
    \includegraphics[width=8.5cm]{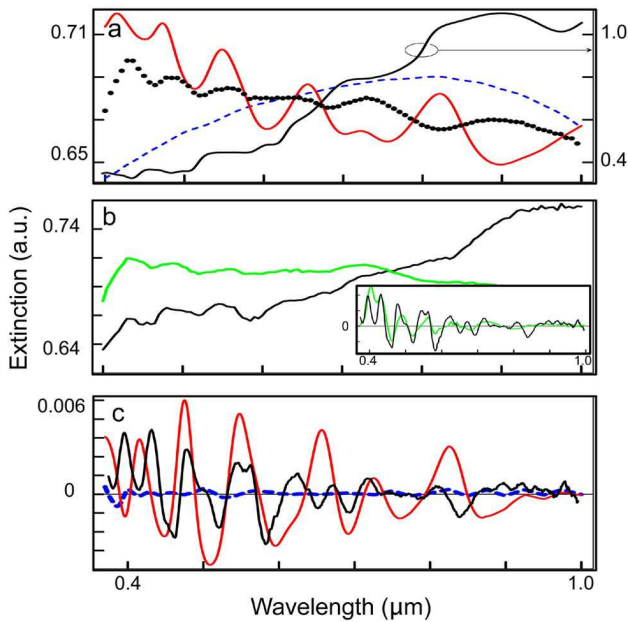}
       \caption{ (a) Comparison of normalized, averaged EMT (blue), and ST (red) extinctions with measured spectra (dots).  Also shown is the extinction obtained using the tabulated bulk silver dielectric function (black). (b) Normalized extinction spectra of two typical thin-shell samples with same core/shell size. Inset: Spectra after subtracting computed background functions.
(c) Comparison of background-subtracted, averaged extinction for EMT (blue), ST (red) and a typical data set (black).  All samples had 1$\mu$m diameter cores and 20nm-thick shells.}
\label{fig:seccomp}
\end{figure}

Nucleation and aggregation of silver crystals onto the spheres result in an intermediate regime of discontinuous nanoshells of thickness 20 - 30nm. Extinction spectra of the spheres in Fig.\ref{fig:sem}(a) reveal several distinct resonance peaks, all shifted from the single-particle plasmon resonance, as shown in Fig.\ref{fig:seccomp}(a). These are core/shell multipole modes, typically observed in high-contrast layered spherical particles. In addition, the inhomogeneously broadened enhanced red extinction tail present in fractal metal aggregates \cite{devaty} is strongly modulated here. These indicate cluster-localized plasmon modes are interacting via cavity Mie resonances. Furthermore, varying the core size affects both peak number and positions, verifying the geometric nature of the resonances. In this new regime of coherently driven localized plasmons collective response of a strongly disordered shell is thus achieved.

The often-used AK formalism, which accounts for multiple scattering in layered concentric shells \cite{aden}, requires knowledge of the dielectric functions of core, shell and embedding medium materials. Using tabulated values for the dielectric function of bulk silver to model the extinction coefficient exhibited serious discrepancies and failed to reproduce the experimental data, as seen in Fig. \ref{fig:seccomp}(a).

As previously indicated, thin-shell samples are insulating, with filling fractions close to the percolation threshold. Hence, we develop a method for modelling the optical response of such quasi-shells based on scaling theory (ST) of percolating planar metal films. The details of ST are described elsewhere \cite{yagil92}, and we summarize only its germane points. Scaling properties of fractal networks of conductors allow the definition of a scaling function \cite{straley}. A dielectric function $\epsilon(\omega;L_{\xi})$ is obtained in terms of this function and a characteristic length scale $L_{\xi}(\omega)$, describing the largest length scale over which homogeneous averaging may occur. $L_{\xi}(\omega)$ is determined by the smallest relevant length scale: the percolation correlation length, $\xi$, or the optical coherence length $L(\omega)$, given by
\begin{subequations}
\begin{eqnarray}
\xi &\propto&  \left(| p - p_c |/p_c\right)^{-\nu}\\
L(\omega) &\propto& \omega^{-1/(2+\theta)}
\end{eqnarray}
\end{subequations}
For a 2D system near percolation the critical exponent is $\nu = 4/3$, and $\theta = 0.79$. A dielectric function $\epsilon_i$ ($\epsilon_c$), corresponding to the thin-shell insulating (conducting) component \cite{yagil92}, is calculated for each length scale $L_{\xi}(\omega)$ using parameters relevant to our system \cite{stparms}.

The main difference between near-percolation planar films and our nanoshells stems from the closed shell geometry, which introduces correlations between film components, otherwise uncorrelated in an open, planar configuration. This allows coherent driving of cluster-localized plasmon resonances through multipole electromagnetic cavity-modes, clearly resolved in the extinction spectra. We account for this cluster-coupling by defining the shell-averaged dielectric function
\begin{equation}
\epsilon_{\rm shell}(\omega) = \langle \epsilon(\omega;L_{\xi}) \rangle = p
\epsilon_c(\omega,L_{\xi}) + (1-p)\epsilon_i(\omega,L_{\xi})
\label{eq:epsshell}
\end{equation}
which is then utilized to calculate the extinction cross section in the AK formalism.

The resulting effective dielectric function exhibits characteristics strikingly different from those of both bulk and nanocrystalline silver. Most importantly, in the visible and near infrared (IR) spectral range Eq. (\ref{eq:epsshell}) yields positive values for Re[$\epsilon_{\rm shell}(\omega)$] and relatively large values for Im[$\epsilon_{\rm shell}(\omega)$], implying a \emph{lossy dielectric response}. This is typical of disordered 2D metals below the percolation threshold. In the IR, Re[$\epsilon_{\rm shell}(\omega)$] crosses over to negative values. For common embedding dielectrics the crossover to metallic behavior occurs at $\lambda\sim1.5-2\mu$m.  This is related to a crossover from inhomogeneous to homogeneous optical response, calculated to occur near $\lambda_\xi\sim650$nm in our samples \cite{yagil92}. For $\lambda\gg\lambda_\xi$ the shell may be treated as optically homogeneous. The consequences of this are discussed in a later section.

The physical properties of composite materials are often treated using an effective-medium approach. According to this method, when spatial inhomogeneities manifest on scales much smaller than the relevant length scale in the system, here the optical wavelength, average homogeneous treatment is justified. However, close enough to the percolation threshold this is known to break down \cite{stroud, yagil92, shalaev2}. We have examined Bruggeman effective medium theory (EMT), best suited for systems near percolation \cite{bruggeman, cohen}, as a plausible model for the shells' response. The effective dielectric function, $\epsilon_{\hbox{\scriptsize eff}}$ is given by
\begin{equation}
p\frac{\epsilon_m-\epsilon_{\hbox{\scriptsize
eff}}}{g\epsilon_m+(1-g)\epsilon_{\hbox{\scriptsize eff}}}
+(1-p)\frac{\epsilon_d-\epsilon_{\hbox{\scriptsize
eff}}}{g\epsilon_d+(1-g)\epsilon_{\hbox{\scriptsize eff}}}=0
\end{equation}
where $\epsilon_m$ and $\epsilon_d$ are known dielectric functions of the interpenetrating metal and the dielectric, respectively, and $p_c=g=0.68$ is the depolarization factor in 2D \cite{yagil92}. Using measured values of $p$ we obtain an expression for $\epsilon_{\hbox{\scriptsize eff}}$, which is then utilized to calculate the extinction cross section as described above. For the relevant experimental parameters \cite{stparms}, $\epsilon_{\hbox{\scriptsize eff}}$ is a positive function over the entire EM spectral range, implying non-metallic response. As a final step for both ST and EMT approaches, the calculated extinction cross sections are averaged over the measured filling fraction $p$ and sphere size distributions \cite{brparms}, and are plotted together for comparison in Fig.\ref{fig:seccomp}(a).

Comparing the models reveals discrepancies in the general trend of the functions, with ST better following the data. Most importantly, the number of resonances, their position and especially contrast, are poorly described by EMT. We note data collected from different batches may exhibit opposite trends, as illustrated by the two typical examples in Fig.\ref{fig:seccomp}(b). This is attributed mainly to various degrees of nanocrystal aggregation present in each batch, where increased aggregation results in a broad, enhanced red absorption tail.  To account for this we first construct a background continuum function by applying a 50nm boxcar average to the extinction spectra. Subtraction of each continuum function from its respective spectrum reveals excellent agreement between the prominent spectral features, shown in Fig.\ref{fig:seccomp}(b) (inset).

Fig.\ref{fig:seccomp}(c) shows background-subtracted spectra, which may be further analyzed using bandpass spectrophotometry. Due to the large number of parameters used by ST as well as the highly complex nature of the multipolar vector spherical harmonics, conventional multivariate fitting methods proved inefficient. Nevertheless, ST clearly best reproduces the experimental data, while EMT lacks the contrast to account for the observed resonances. We attribute the remaining discrepancies between ST and the data to a non-ideal experimental system, its main limitation being irregular shell boundaries, whose additional losses cannot be accounted for within the AK formalism.

As shell thickness increases through deposition of additional silver, we observe shifting of the resonances as in Fig.\ref{fig:sparse}(b). High-resolution SEM micrographs indicate the increase in thickness is accompanied by filling of interparticle voids, as well as shell-coarsening. Shells thicker than 50nm are therefore better treated as 3D aggregates of nanocrystals, with filling fractions estimated at $0.8\lesssim p \lesssim 0.9$. Average measured polydispersities of such shells may still be as low as 5$\%$. Extinction spectra reveal characteristics closely similar to those of continuous silver nanoshells. We may therefore describe the effective dielectric response of porous shells using the classical MG model. The dielectric function $\epsilon_{\hbox{\tiny MG}}$ is given now by
\begin{equation}
\epsilon_{\hbox{\tiny MG}}=\epsilon_m\left[1+\frac{3(1-p)(\epsilon_d-\epsilon_m)/(\epsilon_d+2\epsilon_m)}{1-(1-p)(\epsilon_d-\epsilon_m)/(\epsilon_d+2\epsilon_m)}\right]
\end{equation}
describing a solid silver shell with $1-p$ volume fraction of spherical dielectric inclusions. Following our procedure for calculating the size-averaged extinction cross section \cite{mgparms}, we obtain the plot in Fig.\ref{fig:sparse}(b). We find this model reproduces well our experimental observations.

\begin{figure}[t]
{
  \includegraphics[width=8.5cm]{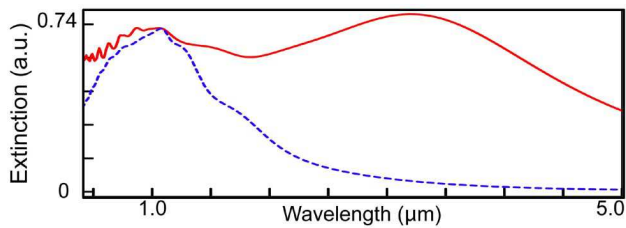}
}
\caption{(Color online) Calculated extinction for ST (line) and EMT (dashed) for $p=0.55$, 20nm--thick silver shells on $1\mu$m diameter silica cores. }
\label{fig:predictedIR}
\end{figure}

An interesting discrepancy between ST and EMT emerges when the extinction calculations are extended to the IR, as in Fig.\ref{fig:predictedIR}. We find ST predicts broad IR absorption, while EMT exhibits a decaying Rayleigh tail. The broad peak near $3.2\mu$m is a mostly dipolar shell plasmon resonance, verified by numerical mode-decomposition of absorption and scattering cross-sections. This is a consequence of the above mentioned ST-crossover to metallic behavior, which enables the excitation of a \emph{delocalized} shell plasmon. Note this IR resonance should be distinguished from anomalous IR absorption in metal-dielectric composites \cite{devaty}. The latter originates from enhanced absorption in the embedding dielectric, due to intense local electric fields \cite{gadenne}. Here we assume both silica core and dielectric inclusions in the shell are lossless, therefore all absorption is due to intrinsic losses in the metal shell.

In summary, we have shown that extinction properties of semicontinuous silver nanoshells grown on dielectric cores unambiguously differ from previously reported continuous core/shell systems. We have demonstrated tuning of the plasmon resonance with shell growth -- from a single dipolar mode in sparse coatings to multiple high order shell excitations in densely packed thick shells. We have identified an intermediate regime of thin fractal shells near the percolation threshold, where geometrically tunable plasmon resonances persist despite disorder and the shells' discontinuous character. A modified approach for modelling the dielectric response of fractal shells, based on scaling theory is consistent with experimental observations. We attribute this new resonant regime to cluster-localized plasmons, coherently driven by micro-resonator modes. Future work will address the predicted delocalized plasmon resonance, resulting from a crossover to homogeneous metallic response in the IR .

The authors thank J. Bouwman and S. Emmons for assistance with chemical syntheses. We also acknowledge helpful discussions with R. Haydock and G. Bothun. This work was supported by NSF Grant No. DMR-02-39273 and ARO Grant No. DAAD19-02-1-0286.



\begin{thebibliography}{25}
\bibitem[{stp()}]{stparms}
\bibinfo{note}{The parameters of the scaling theory model follow \cite{yagil92}
  with the following exceptions: $\sigma_{\hbox{\tiny dc}} = 2.574\times
  10^{17}$sec$^{-1}$, $b_\tau = 3\times10^{-16}$sec, $\xi_0=10$nm, $L_0=1\mu$m
  and $C_0=0.884$. Averaging parameters are the same as in \cite{brparms}.}
\bibitem[{mgp()}]{mgparms}
\bibinfo{note}{Here we fit a single parameter $p$ to the extinction data, using
  the measured value $\bar{T}=70$nm, and obtain $\bar{p}=0.86$.}
\end{thebibliography}
\end{document}